\documentclass[aps,showpacs,preprintnumbers,amsmath, amssymb]{revtex4}

\usepackage{float}
\usepackage{graphics,epsfig}
\usepackage{graphicx}
\usepackage{dcolumn}
\usepackage{bm}

\begin{document}
\baselineskip=0.8 cm

\title{{\bf Analysis of minimum orbital periods around d-dimensional charged black holes }}
\author{Yan Peng$^{1}$\footnote{Corresponding author: yanpengphy@163.com}}
\author{Guohua Liu$^{1,2}$\footnote{Corresponding author: liuguohua1234@163.com}}
\affiliation{\\$^{1}$ College of Mathematics and Physics, Xinjiang Agricultural University, Urumqi, Xinjiang, 830052, China}
\affiliation{\\$^{2}$ School of Mathematical Sciences, Qufu Normal University, Qufu, Shandong 273165, China}

\vspace*{0.2cm}
\begin{abstract}
\baselineskip=0.6 cm
\begin{center}
{\bf Abstract}
\end{center}

This paper investigates the bounds on the minimum orbital period for test objects
around d-dimensional charged black holes in asymptotically flat spacetimes.
We derive the exact critical radius and the
minimum orbital period. We then prove analytically
that the minimum orbital period decreases strictly as the charge of the black
hole increases. Thus, the upper limit is reached for an uncharged black hole,
while the lower limit is attained for a maximally charged one, and the two
bounds take the closed form
$\frac{2\pi(d-2)}{d-3}[(d-2)M]^{\frac{1}{d-3}}\leqslant T_{min}
\leqslant 2\pi\sqrt{\frac{d-1}{d-3}}\,[(d-1)M]^{\frac{1}{d-3}}$.
Since the minimum period equals $2\pi$ times the shadow radius, the upper
bound is equivalently a universal upper bound on the shadow radius.
These results improve our understanding of dynamics around
d-dimensional black holes and impose constraints on candidate gravity theories.

\end{abstract}

\maketitle

\newpage
\vspace*{0.2cm}

\section{Introduction}

The investigation of black holes advances our understanding of
fundamental physical laws and the nature of spacetime.
The motion of objects in curved backgrounds helps us to explore properties of black holes and potential quantum
gravity effects in the extremely curved spacetime \cite{MS1}-\cite{MS12}.
We note that the minimum period is realized by light-like circular orbits,
and for static spherical black holes it equals $2\pi$ times the shadow
radius \cite{MS13}. So the bounds can equivalently be read as
bounds on the shadow radius.
While extensive research has been carried out in 4-dimensional
spacetimes, studies in higher
dimensions are motivated by fundamental theories that predict extra spatial dimensions \cite{ST1,ST2,ST3,ST4}.
Thus, it is meaningful to study the dynamics of test objects moving
around higher-dimensional black holes.

It has been shown that the shortest possible orbital
period is achieved on geodesics in the black hole spacetime \cite{SH1,SHF,YPT}.
In 4-dimensional Kerr black holes, Hod analytically proved
that the minimum orbital period is bounded from below by
$T_{min}\geqslant 4\pi M$, where $T_{min}$ is the minimum
orbital period measured by observers at infinity and $M$ is the
mass of the central compact object \cite{SH1}.
This lower bound was further examined in Kerr-Newman
black hole spacetime, where the lower bound is
still $T_{min}\geqslant 4\pi M$ exactly consistent
with Hod's conjecture \cite{YP1}.
On the other hand, the minimum orbital period of black holes is also bounded from above
by $T_{min}\leqslant 6\sqrt{3}\pi M$ for the Schwarzschild, Reissner-Nordstr\"{o}m, Kerr-Newman,
Hayward and Bardeen black holes \cite{YP2}.
It is therefore meaningful to examine whether similar bounds hold
for the minimum orbital period in more general d-dimensional black holes.

This work studies minimum orbital period bounds of
test objects moving around d-dimensional charged black holes.
We first derive precise expressions of the minimum orbital period.
Then we examine effects of black hole charge on the minimum orbital period
and obtain the upper and lower bounds on the minimum orbital period
through analytical methods.
We present our findings in the final section.

\section{Analytical bounds on minimum orbital period of higher dimensional black holes}

The spacetime serves as an important foundation for understanding the
dynamical behavior. In particular, the black hole metric plays a crucial
role in determining the orbital period of test objects.
The metric of an asymptotically flat, spherically
symmetric, d-dimensional charged black hole reads \cite{HB1}
\begin{equation}
ds^2 = -V(r)dt^2 + \frac{1}{V(r)}dr^2 + r^2 d\Omega_{d-2}^2,
\label{metric}
\end{equation}
with $V(r) = 1 - \frac{2M}{r^{d-3}} + \frac{Q^2}{r^{2(d-3)}}$.
The black hole event horizon is $r=r_{h}$ satisfying $V(r)=0$.
The parameters $M$ and $Q$ are related to the ADM mass $\tilde{M}$ and
electric charge $\tilde{Q}$ through the equations
\begin{equation}
\tilde{M} = \frac{(d-2)\,\omega_{d-2}\,M}{8\pi G},
\end{equation}
\begin{equation}
\tilde{Q} = \frac{\sqrt{2(d-2)(d-3)}\;\omega_{d-2}\,Q}{8\pi G},
\end{equation}
where $\omega_{d-2}$ is the volume of a unit $(d-2)$-sphere.
The angular part $d\Omega_{d-2}^2$ denotes the metric on a (d-2)-dimensional unit sphere.
This part of the metric ensures that the spacetime is
rotationally symmetric in the angular coordinates, typically denoted as
$\theta_1$, $\theta_2$, \ldots, $\theta_{d-3}$, $\phi$. The coordinates
$\theta_i$ (for $i = 1, 2, \ldots, d-3$) are polar angles
ranging from 0 to $\pi$, while $\phi$ is the azimuthal angle
ranging from 0 to $2\pi$.

To study the orbital motion, we consider objects moving along circular paths around such black holes.
For a trajectory to be circular, the radial coordinate $r$ and the angular
coordinates $\theta_1$, $\theta_2$, \ldots, $\theta_{d-3}$ must remain constant.
This means that the differentials of these coordinates must be zero, i.e.,
$dr = d\theta_1 = d\theta_2 = \cdots = d\theta_{d-3} = 0$.
This condition ensures that the object is moving along a circular
orbit without any radial or angular deviation.
To simplify the analysis, we assume that the test objects travel
within the equatorial plane. In this case, the angular coordinates
are fixed at $\theta_1 = \theta_2 = \cdots = \theta_{d-3} = \frac{\pi}{2}$.
Under these conditions, the metric describing the black hole
spacetime can be significantly simplified into
\begin{equation}
ds^2 = - \left(1 - \frac{2M}{r^{d-3}} + \frac{Q^2}{r^{2(d-3)}} \right) dt^2 + r^2 d\phi^2.
\end{equation}

This study aims to determine the minimum orbital period and its
functional dependence on the black hole mass, charge and spacetime dimensionality.
To explore this, we focus on objects approaching the
speed of light. In the context of general relativity, the light speed condition
implies that the spacetime interval $ds^2$ is zero.
Therefore, in the light-speed limit, we obtain the following relation
\begin{equation}
- \left(1 - \frac{2M}{r^{d-3}} + \frac{Q^2}{r^{2(d-3)}} \right) dt^2 + r^2 d\phi^2 = 0.
\end{equation}

A complete revolution corresponds to $d\phi = 2\pi$,
hence we set $d\phi = 2\pi$ to represent a full orbital period.
We also let $dt = T(r)$ denote the time period observed from infinity.
The equation $T(r)$ then satisfies
\begin{equation}
- \left(1 - \frac{2M}{r^{d-3}} + \frac{Q^2}{r^{2(d-3)}} \right) T(r)^2 + r^2 (2\pi)^2 = 0.
\end{equation}
Solving this equation, we obtain the orbital period:
\begin{equation}
T(r) = \frac{2\pi r}{\sqrt{1 - \frac{2M}{r^{d-3}} + \frac{Q^2}{r^{2(d-3)}}}}
     = \frac{2\pi r}{\sqrt{V(r)}}.
\label{Tr}
\end{equation}
This expression for $T(r)$ facilitates the analysis of the orbital
period's dependence on the radius $r$, mass parameter $M$ and charge parameter $Q$.
It provides the foundation for further
investigation into the minimum orbital period and its dependence on these parameters.

To locate the minimum, we differentiate $T$ with respect to $r$ and set the derivative to be
\begin{equation}
\frac{dT}{dr}=\frac{\pi\left(2V-rV'\right)}{V^{3/2}}=0.
\label{dTdr}
\end{equation}
Since $V(r)>0$ outside the horizon, the critical radius is determined by $2V-rV'=0$.
A direct computation gives
\begin{equation}
2V-rV'=2\left[(d-2)Q^{2}r^{-2(d-3)}-(d-1)Mr^{-(d-3)}+1\right].
\end{equation}
The different powers of $r$ in this expression are absorbed by the substitution
\begin{equation}
x=r^{d-3}.
\end{equation}
Multiplying $2V-rV'=0$ by $r^{2(d-3)}/2$, one obtains a quadratic equation in $x$,
\begin{equation}
x^{2}-(d-1)Mx+(d-2)Q^{2}=0,
\label{quadratic}
\end{equation}
with two roots
\begin{equation}
x_{\pm}=\frac{(d-1)M\pm\sqrt{(d-1)^{2}M^{2}-4(d-2)Q^{2}}}{2}.
\label{roots}
\end{equation}

The physical range of the charge is fixed by the horizon.
The horizon equation $V(r_{h})=0$, i.e.
$r_{h}^{2(d-3)}-2Mr_{h}^{d-3}+Q^{2}=0$, is solved by
\begin{equation}
r_{h}^{d-3}=M+\sqrt{M^{2}-Q^{2}},
\end{equation}
so that a genuine black hole requires
\begin{equation}
Q\leqslant M \qquad (d\geqslant 4).
\label{physical}
\end{equation}
Within this whole physical range the discriminant in (\ref{roots}) obeys
\begin{equation}
(d-1)^{2}M^{2}-4(d-2)Q^{2}
\geqslant\left[(d-1)^{2}-4(d-2)\right]M^{2}
=(d-3)^{2}M^{2}\geqslant 0,
\end{equation}
where the first inequality uses $Q^{2}\leqslant M^{2}$.
Hence both roots are real for every allowed charge.

We now show analytically which root lies outside the horizon.
The outer horizon is characterized by $x_{h}=r_{h}^{d-3}=M+\sqrt{M^{2}-Q^{2}}$,
so that $M\leqslant x_{h}\leqslant 2M$.
For the small root, $x_{-}\leqslant M$ is equivalent to
$\sqrt{(d-1)^{2}M^{2}-4(d-2)Q^{2}}\geqslant(d-3)M$, which after squaring reduces to
$4(d-2)(M^{2}-Q^{2})\geqslant 0$, and this holds precisely in the regime (\ref{physical}).
Therefore
\begin{equation}
x_{-}\leqslant M\leqslant x_{h},
\end{equation}
i.e. the small root lies inside the horizon. In this work,
we are only interested in the more physical region outside
the black hole horizon.
For the large root, using again the inequality above,
\begin{equation}
x_{+}\geqslant\frac{(d-1)M+(d-3)M}{2}=(d-2)M\geqslant 2M\geqslant x_{h},
\end{equation}
where the middle inequality uses $d\geqslant4$. Equality in the first estimate
requires $Q=M$ while equality in the last one requires $Q=0$, so they cannot
hold simultaneously, and $x_{+}>x_{h}$ strictly.
Moreover, since $2V-rV'$ is an upward parabola in $x$ with two roots $x_{-}<x_{+}$,
it is negative between the roots and positive for $x>x_{+}$; recalling that
$T(r)\rightarrow\infty$ as $r\rightarrow r_{h}^{+}$ and as $r\rightarrow\infty$,
we conclude that $T(r)$ decreases from the horizon outwards and then increases, so that
\begin{equation}
r_{ph}=x_{+}^{\frac{1}{d-3}}
\end{equation}
is the unique critical point and the global minimum of the orbital period.

It remains to simplify $V$ at the critical point. Using the relation
$(d-2)Q^{2}=(d-1)Mx-x^{2}$ from (\ref{quadratic}), the numerator of $V$ becomes
\begin{equation}
x^{2}-2Mx+Q^{2}
=\frac{(d-2)x^{2}-2(d-2)Mx+(d-1)Mx-x^{2}}{d-2}
=\frac{(d-3)x(x-M)}{d-2},
\end{equation}
and therefore
\begin{equation}
V(r_{ph})=\frac{(d-3)(x-M)}{(d-2)x}.
\label{Vph}
\end{equation}
Since $x=x_{+}\geqslant(d-2)M>M$, one has $V(r_{ph})>0$, as it should be.
Substituting (\ref{Vph}) into (\ref{Tr}), the minimum orbital period takes the closed form
\begin{equation}
T_{min}=2\pi\,x^{\frac{1}{d-3}}
\sqrt{\frac{(d-2)x}{(d-3)(x-M)}},
\qquad x=x_{+}.
\label{Tmin}
\end{equation}

We next examine how $T_{min}$ depends on the charge. Here we prove it
analytically. Regard $T_{min}$ as a function of $x$ along the critical
curve, with $x=x_{+}(Q)$ determined implicitly by (11). Here $M$ and $d$
are fixed parameters, $Q$ is the independent variable, and $x$ is
regarded as a function of $Q$ along the critical curve. Differentiating
the implicit relation (11) with respect to $Q$ via the chain rule, we
obtain
\begin{equation}
2x\frac{dx}{dQ}-(d-1)M\frac{dx}{dQ}+2(d-2)Q=0,
\end{equation}
and hence
\begin{equation}
\frac{dx}{dQ}=-\frac{2(d-2)Q}{2x-(d-1)M}
=-\frac{2(d-2)Q}{\sqrt{(d-1)^{2}M^{2}-4(d-2)Q^{2}}}<0
\qquad (Q>0),
\end{equation}
where the denominator $2x_{+}-(d-1)M$ is exactly the square root of the
discriminant. Indeed, for the larger root (12), a direct substitution
yields
\begin{equation}
2x_{+}-(d-1)M
=\sqrt{(d-1)^{2}M^{2}-4(d-2)Q^{2}}>0.
\end{equation}
Throughout the physical range
$Q\leq M$, the strict inequality holds for every
$d\geq 4$. On the other hand, taking the logarithm of (21) before
differentiating gives
\begin{equation}
\frac{d\ln T_{min}}{dx}
=\frac{1}{(d-3)x}+\frac{1}{2x}-\frac{1}{2(x-M)}
=\frac{2x-(d-1)M}{2(d-3)x(x-M)}>0,
\end{equation}
because $x=x_{+}>(d-1)M/2$ and $x>M$.
Combining the two results,
\begin{equation}
\frac{dT_{min}}{dQ}
=T_{min}\,\frac{d\ln T_{min}}{dx}\,\frac{dx}{dQ}<0
\qquad \text{for } 0<Q\leqslant M,\ d\geqslant 4.
\label{monotone}
\end{equation}
Thus the minimum orbital period decreases strictly as the charge
increases throughout the entire black hole parameter space.

An immediate consequence of (\ref{monotone}) is that the maximum of $T_{min}$ in
the black hole family is attained at $Q=0$ (the neutral black hole),
while the minimum is attained at $Q=M$ (the extremal black hole).
At $Q=0$ one has $x=(d-1)M$ and $V(r_{ph})=\frac{d-3}{d-1}$, so
\begin{equation}
T_{min}\big|_{Q=0}
=2\pi\sqrt{\frac{d-1}{d-3}}\left[(d-1)M\right]^{\frac{1}{d-3}}.
\end{equation}
At $Q=M$ one has $x=(d-2)M$ and $V(r_{ph})=\frac{(d-3)^{2}}{(d-2)^{2}}$, so
\begin{equation}
T_{min}\big|_{Q=M}
=\frac{2\pi(d-2)}{d-3}\left[(d-2)M\right]^{\frac{1}{d-3}}.
\end{equation}
We thus establish rigorous bounds for the minimum orbital period as follows
\begin{equation}
\boxed{\ \frac{2\pi(d-2)}{d-3}\left[(d-2)M\right]^{\frac{1}{d-3}}
\leqslant T_{min}
\leqslant 2\pi\sqrt{\frac{d-1}{d-3}}\left[(d-1)M\right]^{\frac{1}{d-3}}\ }
\label{bounds}
\end{equation}
where the upper limit is obtained in the neutral limit and the lower
limit is saturated at maximal charge.
We also note that $T_{min}=2\pi r_{ph}/\sqrt{V(r_{ph})}=2\pi R_{sh}$,
so the upper bound in (\ref{bounds}) is equivalently a universal upper
bound on the shadow radius of d-dimensional charged black holes,
consistent in spirit with the photon-sphere theorem of \cite{MS13}.
It is worth noting that both bounds in (29) approach the same limit
$2\pi$ as $d\to\infty$ with fixed $M$.
Hence $T_{min}\to 2\pi$, or equivalently $R_{sh}\to 1$.

For $d=5$, the bounds
(\ref{bounds}) degenerate to
\begin{equation}
3\sqrt{3}\,\pi\sqrt{M}\leqslant T_{min}\leqslant 4\sqrt{2}\,\pi\sqrt{M}.
\end{equation}

For completeness, setting $d=4$ in (\ref{bounds}) yields
\begin{equation}
8\pi M\leqslant T_{min}\leqslant 6\sqrt{3}\,\pi M,
\end{equation}
in agreement with
bounds obtained for the four dimensional Reissner-Nordstr\"{o}m black holes \cite{SH1}.

\section{Conclusions}

In this paper, we have studied the bounds on the minimum orbital period
around d-dimensional charged black holes. By applying analytical methods,
we have derived the expressions for the minimum orbital period and analyzed
the influence of charge and spacetime dimensionality on the minimum orbital period.
We proved analytically that the minimum orbital period decreases strictly
as the black hole charge increases, and consequently established the bounds
$\frac{2\pi(d-2)}{d-3}\left[(d-2)M\right]^{\frac{1}{d-3}}
\leqslant T_{min}
\leqslant 2\pi\sqrt{\frac{d-1}{d-3}}\left[(d-1)M\right]^{\frac{1}{d-3}}$,
where the upper limit is saturated by neutral black holes and the lower
limit is attained by extremal black holes.
Since $T_{min}=2\pi R_{sh}$ with
$R_{sh}$ the shadow radius, the upper bound also provides a universal
upper bound on the shadow radii of these black holes.
Moreover, both bounds converge to the same constant $2\pi$ as
$d\to\infty$ (with fixed $M$), so that $T_{min}\to 2\pi$, or
equivalently $R_{sh}\to 1$.
Our results provide valuable insights into the dynamics of particles in
the vicinity of these black holes. They also improve our understanding
of the properties of black holes in higher-dimensional spacetimes.

\begin{acknowledgments}

This work was supported by the Shandong Provincial Natural Science Foundation of China under Grant
No. ZR2022MA074. This work was also supported by a grant from Qufu Normal University
of China under Grant No. xkjjc201906.

\end{acknowledgments}

\end{document}